%% file: SQUEzE.tex
\documentclass[fleqn,usenatbib]{mnras}
\usepackage[utf8]{inputenc}
\usepackage[catalan, english]{babel}
\usepackage{graphicx}
\graphicspath{{./figures/}}
\usepackage{natbib}
\usepackage{appendix}
\usepackage{tabularx}
\usepackage{color}
\usepackage{booktabs}
\usepackage{threeparttable}
\usepackage{longtable}
\usepackage{multicol}
\usepackage{multirow}
\usepackage{enumerate}
\usepackage{amsmath}
\usepackage{amssymb}
\usepackage{setspace} 
\usepackage{hyperref}
\usepackage[english]{cleveref}
\usepackage{float}

\definecolor{halfgray}{gray}{0.55}
\definecolor{naviblue}{RGB}{0,0,102}
\definecolor{webbrown}{rgb}{.6,0,0}
\definecolor{RoyalBlue}{cmyk}{1, 0.50, 0, 0}
\definecolor{webgreen}{rgb}{0,.5,0}
\definecolor{Maroon}{cmyk}{0, 0.87, 0.68, 0.32}
\definecolor{Black}{cmyk}{0, 0, 0, 0}
\definecolor{myorange}{RGB}{239, 186,67}

\hypersetup{%
    colorlinks=false,linktoc=all, pdfstartpage=1, pdfstartview=FitV,%
    breaklinks=true, pdfpagemode=UseNone, pageanchor=true, pdfpagemode=UseOutlines,%
    plainpages=false, bookmarksnumbered, bookmarksopen=true, bookmarksopenlevel=2,%
    hypertexnames=true, pdfhighlight=/O,
    urlcolor=webbrown, linkcolor=RoyalBlue, citecolor=webgreen,
}

\def\aap{A\&A}

\def\apjs{ApJS}
\def\apjl{ApJL}

\crefname{equation}{equation}{equations}
\crefname{figure}{figure}{figures}
\crefname{appsec}{appendix}{appendices}
\creflabelformat{equation}{#2#1#3}

\newcommand{\lyb}{Ly$\beta$/O\thinspace VI}
\newcommand{\lya}{Ly$\alpha$}
\newcommand{\siiv}{Si\thinspace IV}
\newcommand{\civ}{C\thinspace IV}
\newcommand{\civred}{\civ{} (red)}
\newcommand{\civblue}{\civ{} (blue)}
\newcommand{\ciii}{C\thinspace III}
\newcommand{\neiv}{Ne\thinspace IV}
\newcommand{\mgii}{Mg\thinspace II}
\newcommand{\oii}{O\thinspace II}
\newcommand{\nev}{Ne\thinspace V}
\newcommand{\hb}{H$\beta$}
\newcommand{\oiii}{O\thinspace III}
\newcommand{\ha}{H$\alpha$}

\newcommand{\angs}{\textup{\AA}}
\newcommand{\kms}{\thinspace km/s}

\begin{document}

\title[SQUEzE I: Methodology]{Spectroscopic QUasar Extractor and redshift (z) Estimator SQUEzE I: Methodology}
\date{}
\input{src/authors.tex}

\maketitle

\begin{abstract}
  \input{src/SQUEzE_abstract.tex}
\end{abstract}


\input{src/SQUEzE_introduction.tex}
\input{src/SQUEzE_squeze.tex}
\input{src/SQUEzE_data.tex}
\input{src/SQUEzE_convergence_tests.tex}
\input{src/SQUEzE_results.tex}

\input{src/SQUEzE_discussion.tex}
\input{src/SQUEzE_conclusions.tex}

\input{src/acknowledgments.tex}

\bibliographystyle{apj}
\bibliography{iprafols}

\appendix
\input{src/SQUEzE_optimization_step1.tex}

\end{document}

%% file: src/authors.tex
\author[Ignasi P\'erez-R\`afols et al.]
  {Ignasi ~P\'erez-R\`afols,$^{1}$\thanks{email: ignasi.perez@lam.fr}, Matthew M. Pieri$^{1}$, Michael ~Blomqvist$^{1}$,
\newauthor Sean ~Morrison$^{1}$, Debopam ~Som$^{1}$
\\
$^{1}$Aix Marseille Univ, CNRS, CNES, LAM, Marseille, France\\
}

%% file: src/SQUEzE_abstract.tex
We present SQUEzE, a software package to classify quasar spectra and estimate their redshifts. SQUEzE is a random forest classifier operating on the parameters of candidate emission peaks identified in the spectra. We test the performance of the algorithm using visually inspected data from BOSS as a truth table. Only 4\% of the sample ($\sim$6,800 quasars and $\sim$11,520 contaminants) is needed for converged training in recommended choices of the confidence threshold ($0.2<p_{\rm min}<0.7$). For an operational mode which balances purity and completeness  ($p_{\rm min}= 0.32$) we recover a purity of $ 97.40\pm0.47\%$ ($ 99.59\pm0.06\%$ for quasars with $z \geq 2.1$) and a completeness of $ 97.46\pm0.33\%$ ($ 98.81\pm0.13\%$ for quasars with $z \geq 2.1$). SQUEzE can be used to obtain a $\approx$100\% pure sample of  $z \geq 2.1$. quasars (with $\approx$97\% completeness) by using a confidence threshold of $p_{\rm min}=0.7$. The estimated redshift error is $1,500{\rm \kms}$ and  we recommend that SQUEzE be used in conjunction with an additional step of redshift tuning to achieve maximum precision. We find that SQUEzE achieves the necessary performance to replace visual inspection in BOSS-like spectroscopic surveys of quasars with subsequent publications in this series exploring expectations for future surveys and alternative methods.


%% file: src/SQUEzE_introduction.tex
\section{Introduction}\label{sec:introduction}

In recent years observational cosmology has undergone a substantial change. We have moved from relatively small datasets that allowed us to indicatively measure cosmological model parameters, to huge datasets that have allowed us to constrain these parameters to the percent level. We have entered the so-called precision cosmology era. Surveys such as the Sloan Digital Sky Survey \citep[SDSS;][]{York+2000, Abazajian+2009}, the 2-degree Field Galaxy Redshift Survey \citep[2dFGRS;][]{Colless+2001}, the Wilkinson Microwave Anisotropy Probe \citep[WMAP;][]{Bennett+2003}, the 6-degree Field Galaxy Survey \citep[6dFGS;][]{Jones+2004}, the WiggleZ survey\citep{Drinkwater+2010}, Planck \citep{PlanckI2011}, the Baryon Oscillation Spectroscopic Survey \citep[BOSS;][part of the SDSS-III survey \citealt{Eisenstein+2011}]{Dawson+2013}, the extended BOSS \citep[eBOSS;][part of SDSS-IV \citealt{Blanton+2017}]{Dawson+2016}, or Gaia \citep{Gaia+2016}, amongst others, have created huge datasets. These datasets are not only photometric catalogues, but often are comprised of hundreds of thousands of spectra, and are expected to become even larger in the next generation of surveys: WEAVE-QSO (\citealt{Pieri+2016}; as part of WEAVE, \citealt{Dalton+2016}), DESI \citep{DESI2016}, Euclid \citep{Laureijs+2010}, 4MOST \citep{deJong+2016}, WFIRST \citep{Dressler+2012,Green+2012,Spergel+2015}, PSF \citep{Takada+2014,Chiba+2016}, ...

The generation of such large datasets of spectra has not only allowed us to improve the constraints on the existing models, but have enabled other probes to be explored. Indeed, we can now measure the baryon acoustic oscillation (BAO) peak from the distribution of galaxies \citep[e.g.][]{Anderson+2012, Anderson+2014a, Anderson+2014b, Tojeiro+2014, Bautista+2017b} at the percent level precision. In particular, we have seen the birth of the \lya{} cosmology \citep[e.g.][]{McDonald+2006, Slosar+2011, Busca+2013}, which exploits the almost continuous `forest' of \lya{} absorption bluewards of the \lya{} emission line in quasar spectra. Many things have been achieved using the \lya{} forest, including measurements of BAO at $z\sim2.25$ \citep{Bautista+2017, duMasdesBourboux+2017}, preliminary measurements of BAO at lower redshifts using metal absorption \citep{Blomqvist+2018, duMasdesBourboux+2019}, and constraints on the neutrino mass using the one-dimensional power spectrum \citep[e.g.][]{Palanque-Delabrouille+2013}.

In addition to observational cosmology, these large datasets have also allowed for new studies of the Intergalactic Medium (IGM), focusing on average properties of the gas. Stacking techniques allow us to measure average properties of IGM even in moderate signal-to-noise spectra. Examples of these measurements include the characterization of the average spectrum of the Damped \lya{} Absorbers \citep[DLAs;][]{Mas-Ribas+2017}, their metal strength \citep{Arinyo-i-Prats+2018}, and bias factor \citep{Perez-Rafols+2018,Perez-Rafols+2018b}. Additionally, stacking techniques can also help us probe the Circumgalactic Medium (CGM), as is the case with the strong \lya{} abosorbers \citep{Pieri+2014, Som+InPrep}. 

In this context, correctly identifying spectra of quasars plays a key role in the success of these (and other) analysis. Having non-quasars classified as quasars, or classifying them with a completely wrong redshift, dilutes any of the measurements due to the addition of noise to the sample. In surveys such as BOSS \citep{Dawson+2013}, from the SDSS-III collaboration \citep{Gunn+1998, York+2000, Gunn+2006, Eisenstein+2011, Bolton+2012, Smee+2013}, a visual inspection of all things targeted as quasars was performed, which resulted in a quasar catalogue, the latest version of which being the DR12Q catalogue \citep{Paris+2017}. This solution will clearly become infeasible as datasets grow larger and larger in the next generation of surveys.

We present a fully automated machine learning code in \texttt{Python}, SQUEzE, to identify quasars from a set of spectra and give a rough estimate of its redshift. The main objective of this code is to exploit the self-similarity of quasars in order to identify them, in a sense following the steps performed in the visual inspection in a fully automated way. Additionally, the code can flexibly ingest spectra of different resolution and is light both in training as well as operation mode. The code and the methodology and simple performance tests are presented in this paper, and we will present a more detailed study of SQUEzE performance and the robustness of its results in follow-up papers. 

This article is structured as follows. In Section~\ref{sec:squeze} we present the code and explain its algorithm. The code is tested by using a visually inspected quasar sample, described in Section~\ref{sec:sample}. In Section~\ref{sec:convergence} we perform convergence tests to determine the amount of objects required to construct sufficiently large samples. The performance of the code is presented in Section~\ref{sec:results}, but we stress that a more detailed study of this performance will be given in follow-up papers. Finally we discuss these results in Section~\ref{sec:discussion} and conclude in Section~\ref{sec:conclusions}.

%% file: src/SQUEzE_squeze.tex
\section{SQUEzE Overview}\label{sec:squeze}

In this section we present the Spectroscopic QUasar Extractor and redshift (z) Estimator, SQUEzE, a \texttt{Python} code to identify quasars. As stated above, the main purpose of SQUEzE is to automatically identify quasars from a set of given spectra, and give a rough estimate of their redshifts. We will start by presenting a general overview of the code in Section~\ref{sec:squeze_general}, followed by a description of the different modes of operation available in SQUEzE in Section~\ref{sec:squeze_modes}. The code is written in Python and is publicly available at \url{https://github.com/iprafols/SQUEzE}.

\subsection{General overview}\label{sec:squeze_general}

SQUEzE is designed to mimic the process of human visual inspection of the spectra to identify and estimate redshifts of quasars. Broadly speaking, this means going spectrum-by-spectrum through the sample, identifying the position of emission peaks, hypothesising emission line identities to those peaks, checking the placement of other expected emission lines based on that assumption, and deciding which hypothesis, if any, provides a reliable quasar redshift estimate. The presence of emission peaks in a given spectrum does not guarantee that the studied object is a quasar, and the human brain does an effective job of using the combined information of all the emission peaks in the spectrum to decide whether it is, or it is not, a quasar.

In order to reproduce the essential elements of visual inspection in automated form, SQUEzE is organised in three steps performed for each  candidate quasar spectrum:

\begin{enumerate}

\item All significant peaks are found and each one is assigned various metrics associated with quasar emission line identifications. This list of peak-indentification pairs constitutes the set of trial quasar redshifts assessed for that spectrum. For each trial redshift a series of high-level metrics are calculated intended to probe for emission lines that might confirm the trial redshifts.

\item  The set of calculated metrics for each trial redshift are then passed to a trained machine learning classifier (a random forest classifier) in order to quantify the quality of this trial redshift. 

\item A decision is taken on whether any trial quasar redshifts are sufficiently acceptable to merit inclusion in a quasar catalogue and if so, which one is the most reliable. 

\end{enumerate}

We will now review each of these steps in more detail.

\subsubsection{Step 1: trial redshifts}

The first step in SQUEzE is to analyse each of the spectra to look for emission peaks. To achieve this SQUEzE performs a peak finding in a smoothed version of the spectra and keeps only those peaks above a specified significance threshold. The smoothing is done by convolving the flux with a Gaussian kernel of a specified width. The choice for width is 70 pixels, and the significance threshold is set to 6. This choice is intended for BOSS spectra analysed here (see Section~\ref{sec:sample}, below) and can be changed at runtime. The reasoning behind this choice is described in appendix \ref{sec:opt_step1}.

Once we have a list of emission peaks, we append useful information to it, information that will be used in the following steps to classify the objects. We want the code to be light, efficient, applied to many hundreds of thousands of spectra, and avoid tuning to peculiarities of instrumentation or pipeline reduction. Hence SQUEzE does not see the entirety of the spectra in the steps which follow, but only a few key metrics that distill the key information needed. For SQUEzE to reproduce what the human eye does, it is directed at very specific parts of the spectra to verify whether emission line peaks are present. To this end, we define a set of emission lines that will be assessed. For each of the lines SQUEzE computes three metrics: the line ratio, the line contrast ratio, and the local continuum slope. These metrics will be used to decide whether there is a line or not at the specified position. They are defined as follows:
\begin{equation}
	\label{eq:line_ratio}
	\text{line ratio: } l_{i} = \frac{2p_{i}}{b_{i} + r_{i}} ~,
\end{equation}
\begin{equation}
	\label{eq:line_contrast_ratio}
	\text{line contrast ratio: } c_{i} = \left(l_{i}-1\right)/e_{i} ~,
\end{equation}
\begin{equation}
	\label{eq:continuum_slope}
	\text{line continuum slope: } s_{i} = \left|\frac{r_{i} - b_{i}}{r_{i} + b_{i}}\right| ~.
\end{equation}
Here $i$ is a subindex that runs over all the specified lines, $p_{i}$ is the flux average over a peak window, and $b_{i}$ and $r_{i}$ are the flux average over a continuum window on the blue and red side of the line respectively, and $e_{i}$ is the noise estimate of the ratio, computed using standard error propagation. Whenever any of these metrics is not computable (because e.g., the variance is zero, there is missing data, or simply because the selected line is out of range) it is assigned the value \texttt{NaN}.

\begin{figure}
	\centering
	\includegraphics[width=0.45\textwidth]{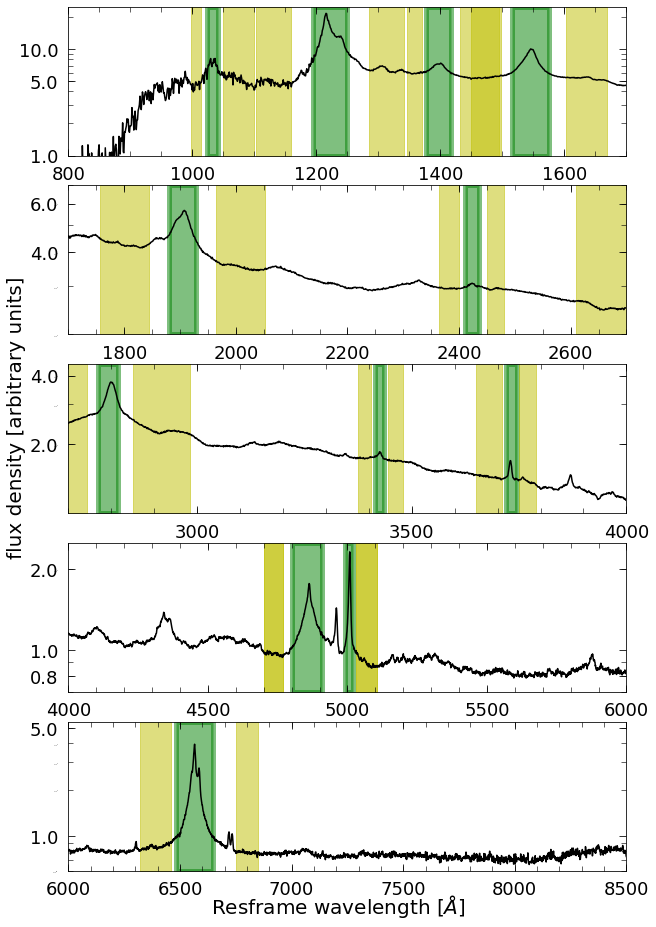}
	\caption{Overplotted to the quasar mean composite from \protect\cite{VandenBerk+2001}, the emission, blue, and red windows for the selected lines. Blue and red windows are plotted in yellow, and darker regions signal overlap between windows in different lines. Emission windows are plotted as green regions with thick lines. \civred{} and \civblue{} lines are not shown in this plot because of their complete overlap with the \civ{} line.}
	\label{fig:lines}
\end{figure}

A summary of the chosen lines, and the peak, blue, and red windows is given in Table~\ref{ta:lines}, and visually displayed in Figure~\ref{fig:lines}. The chosen lines are \lyb{}, \lya{}, \siiv{}, \civ{}, \ciii{}, \mgii{}, \oiii{}, \hb{}, \oii{}, and \ha{}. Additionally, we include \neiv{} and \nev{} which are substantially weaker, but their wavelength are such that they can introduce substantial line confusion if the peaks were detected. We will come back to this point later on. In order to deal with the identification of quasars with Broad Absorption Lines (BALs), we also split the \civ{} line into \civred{} and \civblue{}. The reasoning behind this choice is 
to formulate a metric which returns values that are not diluted by the trough. 

\begin{table*}
	\centering
	\begin{tabular}{cccccccc}
		\toprule
		\midrule
		\multirow{2}{*}{line} & \multirow{2}{*}{wavelength} & \multicolumn{2}{c}{emission window} & \multicolumn{2}{c}{blue window} & \multicolumn{2}{c}{red window} \\
		& & start & end & start & end & start & end \\
		\midrule
		\lyb{} & 1033.03	 & 1023.0 & 1041.0 & 998.0 & 1014.0 & 1050.0 & 1100.0 \\
		\lya{} & 1215.67	 & 1194.0 & 1250.0 & 1103.0 & 1159.0 & 1285.0 & 1341.0 \\
		\siiv{} & 1396.76 & 1377.0 & 1417.0 & 1346.0 & 1370.0 & 1432.0 & 1497.0 \\
		\civ{} & 1549.06 & 1515.0 & 1575.0 & 1449.5 & 1494.5 & 1603.0 & 1668.0 \\
		\civblue{} & 1549.06 & 1515.0 & 1549.06 & 1449.5 & 1494.5 & 1603.0 & 1668.0 \\
		\civred{} & 1549.06 & 1549.06 & 1575.0 & 1449.5 & 1494.5 & 1603.0 & 1668.0 \\
		\ciii{} & 1908.73 & 1880.0 & 1929.0 & 1756.0 & 1845.0 & 1964.0 & 2053.0 \\
		\neiv{} & 2423.83 & 2410.0 & 2435.0 & 2365.0 & 2400.0 & 2450.0 & 2480.0 \\
		\mgii{} & 2798.75 & 2768.0 & 2816.0 & 2610.0 & 2743.0 & 2851.0 & 2984.0 \\
		\oii{} & 3728.48 & 3720.0 & 3745.0 & 3650.0 & 3710.0 & 3750.0 & 3790.0 \\
		\nev{} & 3426.84 & 3415.0 & 3435.0 & 3375.0 & 3405.0 & 3445.0 & 3480.0 \\
		\hb{} & 4862.68 & 4800.0 & 4910.0 & 4700.0 & 4770.0 & 5030.0 & 5105.0 \\
		\oiii{} & 5008.24 & 4990.0 & 5020.0 & 4700.0 & 4770.0 & 5030.0 & 5105.0 \\
		\ha{} & 6564.61 & 6480.0 & 6650.0 & 6320.0 & 6460.0 & 6750.0 & 6850.0 \\
		\bottomrule
	\end{tabular}
	\caption{Summary of the properties of the selected lines. The columns show the line's name, the nominal wavelength of the line, and the starting and ending wavelengths for the emission, blue, and red windows. These windows are used to compute the line ratio, the line contrast ratio and the line continuum slope, as explained in the text. All wavelengths are given in $\angs$ at the restframe.}
	\label{ta:lines}
\end{table*}

In order to translate the restframe wavelengths to observed wavelength SQUEzE assigns several {\it trial redshift}, $z_{\rm try}$, to each of the emission line peaks. These redshifts correspond to assuming that the emission peak corresponds to the \lya{}, \civ{}, \ciii{}, \mgii{}, \hb{}, and \ha{} emission lines (see Table~\ref{ta:lines}). The trial redshift is then computed using both the wavelength of the lines, $\lambda_{l}$ and the detected wavelength for the center of the peak $\lambda_{p}$:
\begin{equation}
	\label{eq:z_try}
	z_{\rm try} = \lambda_{p}/\lambda_{l} - 1 ~.
\end{equation}
Whenever the computed trial redshift is negative, the entry is discarded. Estimates of quasar redshifts based on emission line peak wavelengths vary with respect to the true redshift of the quasar due to their non-negligible bulk velocities \citep[see e.g.][]{Shen+2016} and uncertainty in the peak position measurement. As a result uncorrected redshifts directly estimated from emission lines peak positions suffer from significant systematic and stochastic errors. Hence, these trial redshift are estimates of the true redshift, which can be improved by some additional form of redshift tuning. A discussion on this offset is given later in Section~\ref{sec:results_z}.

\begin{figure}
	\centering
	\includegraphics[width=0.5\textwidth]{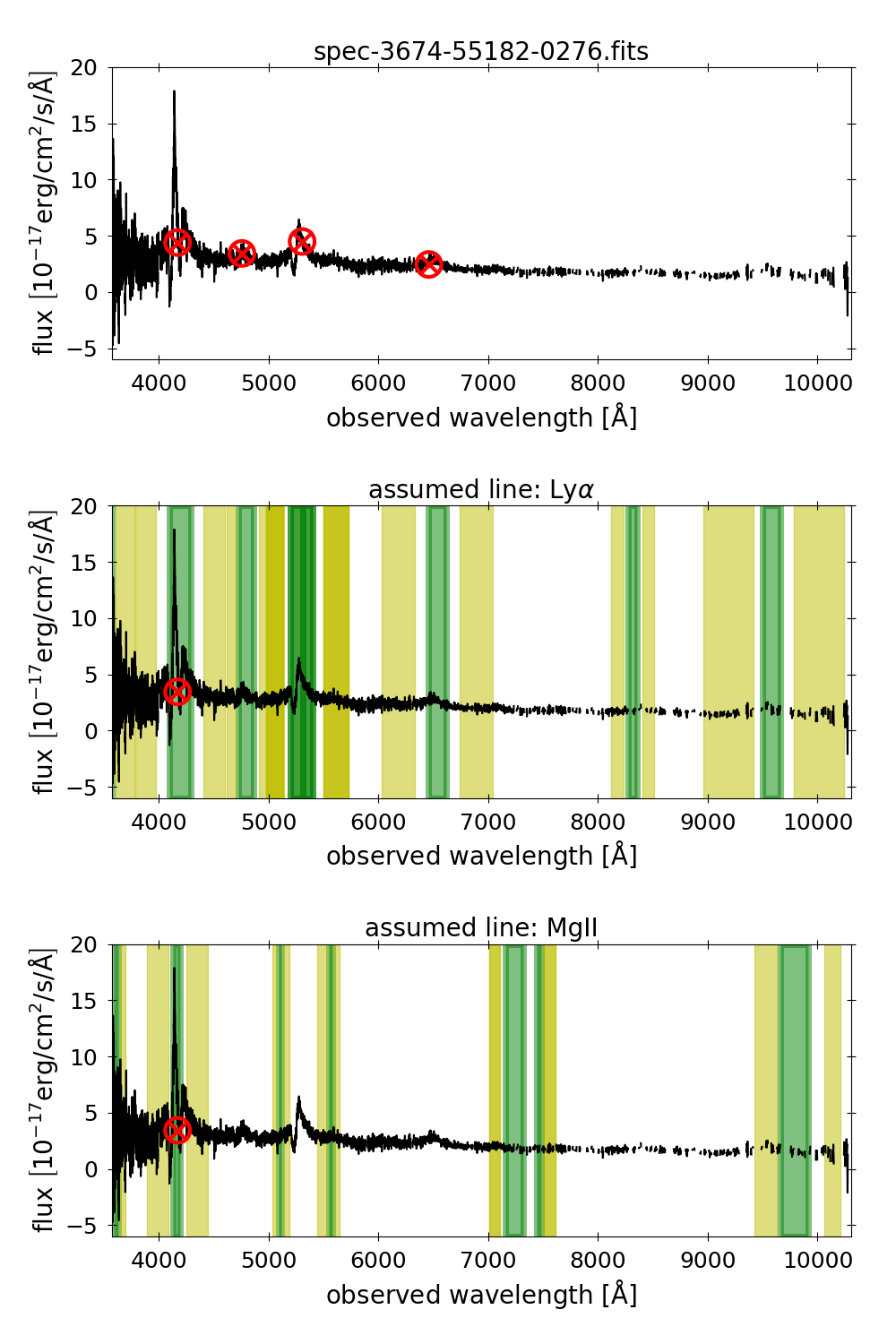}
	\caption{Illustration of step 1. First, peaks are located (top panel, red crossed circles). Then for each of the peaks we compute the peak metrics assuming each peak is each of the selected lines. Middle and bottom panels illustrate the computation of these metrics when the second peak is assumed to be \lya{} (middle panel) and \mgii{} (bottom panel). Blue and red windows are plotted in yellow, and darker regions signal overlap between windows in different lines. Emission windows are plotted as green regions with thick lines. The fact that the peaks are in different positions when assuming different lines will help us filter the list in later steps of the code.}
	\label{fig:step1}
\end{figure}

In summary, after step 1 is completed, SQUEzE will have generated a list of trial redshifts that will contain an object id, the trial redshift, the metrics for all the lines specified in Table~\ref{ta:lines}, and any other catalogue information given for the object that will not be used by SQUEzE. Step 1 is summarised in Figure~\ref{fig:step1}. The top panel illustrates the initial search of the peaks for a sample spectrum, and the middle and bottom panels illustrate the behaviour of the code for a given peak hypothesising that the peak may correspond to several lines.

\subsubsection{Step 2: Classifier}
The second step in the SQUEzE algorithm consists of taking high-level metrics for each trial redshift from the first step
 and assessing them to determine the degree of consistency with expectations of quasar spectra.
 In this step we use a Random Forest classifier, in particular the \texttt{ensemble.RandomForestClassifier} class of the \texttt{Python} module \texttt{sklearn} \citep{sklearn}.
Generically, the Random Forest algorithm fits a number of decision tree classifiers on various sub-samples of the dataset and uses averaging to improve the predictive accuracy and control over-fitting. 

The training of random forests allows the algorithm to learn to identify groups of data based on the input values of the code. In the case of SQUEzE, this means not only learning to separate quasars from stars and galaxies, but also learning to separate from quasars with the correct and incorrect redshifts.
In our training sample, we classify all the entries into several classes. The description and numerical code of these classes is given in Table~\ref{ta:classes}. To determine whether the entry is at the correct redshift we require $\left|z_{\rm try} - z_{\rm true}\right| \le \Delta z_r$. The choice of this redshift precision requirement must be made with care. 
A selected value of  $\Delta z_r$ that is either too small or too large will lead to misclassifications. Both cases introduce noise in the training sample that will reduce the efficiency  of the classifier. We found a value of $\Delta z_r=0.15$ to be satisfactory. We stress, however, that our final redshift precision is much better ($\sim$0.018) than this simple requirement limit, as shown in Section~\ref{sec:results_z}.
  
\begin{table}
	\centering
	\begin{tabular}{lc}
	\toprule 
	Class description & Code\\
	\midrule
	Star & 1\\
	Quasar & 3\\
	Quasar, wrong z & 35\\
	Quasar, BAL & 30\\
	Quasar, BAL, wrong z & 305 \\
	Galaxy & 4\\
	Galaxy, wrong z & 45\\
	\bottomrule
	\end{tabular}
	\caption{Description and numerical code of the different classes.}
	\label{ta:classes}
\end{table}

In general, the Random Forest classifier assigns a probability of an object belonging to each of the used classes. We require a parameter $p$ that reflects our confidence that the trial redshift is approximately correct one for a quasar. SQUEzE computes $p$  for each case as the sum of the probabilities of the object belonging to classes 3 ({\it Quasar}) and 30 ({\it Quasar, BAL}). Because we are only assessing spectra where emission peaks are detected, and because we are only going to keep one entry for each spectrum, we stress that this $p$ value should not be though of as a probability of the entry being a quasar, but rather as a confidence parameter. As shown in Section~\ref{sec:results}, including low values of this confidence parameter yields very good purity and completeness on the final catalogue. 

In practice we split step 2 into two tasks: the identification of quasars with $z_{\rm try} \ge 2.1$ and $z_{\rm try} < 2.1$. This is motivated by the presence or lack (respectively) of  a \lya{} emission line in the spectral coverage, which has a strong impact on the ease of classification. When training, two independent instances of the Random Forest classifiers are trained one for each of the subsamples. When classifying, the $z_{\rm try} \geq 2.1$ ($z_{\rm try} < 2.1$) trial redshifts are built using the classifier trained on the $z_{\rm try} \geq 2.1$ ($z_{\rm try} < 2.1$) sample. Note that this split is optional, but we found that it increased the purity and completeness of the final catalogue.

\subsubsection{Step 3: Final classification}

At this stage, SQUEzE has a list of entries with the following information: trial redshift, the peak metrics for all the lines specified in Table~\ref{ta:lines}, the $p$ value, and any other catalogue information given for the object that will not be used by SQUEzE. However, a given object can appear more than once in this list. For example, a quasar will generally have more than one emission peak, and will be in the list many times. To produce the final catalogue we have to filter all the repeated entries, leaving only one redshift per object. This is done by comparing the $p$ values in the entries of the same object. The entry with maximal $p$ value is kept and the rest are discarded. When constructing a catalogue, SQUEzE offers the possibility to perform a confidence threshold cut. If such a cut is applied, quasars with a lower confidence will be discarded.

\subsection{Modes of operation}\label{sec:squeze_modes}

We have described the SQUEzE algorithm in the previous section and we have seen that SQUEzE uses supervised learning classifiers. The classifiers need to be trained prior to its usage. As such, SQUEzE has different modes of operation. The available modes are {\it training}, {\it test}, {\it operation}, and {\it merge}. We now describe the particularities of each of these modes.
\begin{enumerate}
	\item {\it Training mode}: The training mode is used to train the Random Forest classifiers in step 2. The true identity of the input spectra must be known. In particular, the true redshifts must be given as inputs to training. They are, for example, mock spectra or visually inspected ones. In our demonstration here we use real data with visual inspections. 
	\item {\it Test mode}: The test mode is used to estimate the performance of SQUEzE. The true identity of the sample used in this mode must also be known, and it must be independent of the sample used for training.
	\item {\it Operation mode}: The operation mode is used in unknown data to do the actual classification.
	\item {\it Merge mode}: The merge mode is used to merge the outputs of different runs, in case the input dataset is too large to be run in one go.
\end{enumerate}

%% file: src/SQUEzE_data.tex
\section{Training, validation and test samples}\label{sec:sample}
As explained in Section~\ref{sec:squeze_general} the code uses machine learning techniques to classify the objects as quasars and estimate their redshifts. It is then imperative that we define the training, test, and validation samples. Additionally, we require different train and validation samples to perform the convergence tests in Section~\ref{sec:convergence}. 

To create these samples we use a large, publicly available dataset that contains visually inspected spectra and redshift estimation \citep{Paris+2017}. This sample is based on the objects targeted to be quasars \citep{Ross+2012} in the final Data Release (DR12) of BOSS \citep{Dawson+2013}, from the SDSS-III collaboration \citep{Gunn+1998, York+2000, Gunn+2006, Eisenstein+2011, Bolton+2012, Smee+2013}, and is known as the {\it superset} catalogue\footnote{available at \url{https://data.sdss.org/sas/dr12/boss/qso/DR12Q/Superset_DR12Q.fits}}.  This sample has broadband colours consistent with that of quasars.

This catalogue contains 546,856 objects of which we discard those entries where \texttt{Z\_CONF\_PERSON} is not 3. This cut removes entries in which the visual classification was not firm. We are left with 525,302 objects. We additionally exclude objects selected by ancillary programs, leaving 447,958 objects. Out of these 226,580 are classified as quasars (\texttt{CLASS\_PERSON} = 3), 24,077 as quasars with BAL (\texttt{CLASS\_PERSON} = 30), 182,627 as stars (\texttt{CLASS\_PERSON} = 1), and 14,674 as galaxies (\texttt{CLASS\_PERSON} = 4). 

Now that we have defined the general dataset that we will be using, we explain how the training, test, and validation samples are drawn from it. For simplicity, we define the size of the sample using the number of plates in the sample. We create a set of training and validation samples by randomly drawing plates (without repetition). We start by creating a set containing 2 plates, and we create the subsequent sets of the groups by adding as many plates as the previous set of the group, until we reach a set of 512 plates. We restrict the plates chosen to those belonging to the BOSS survey, which have on average 100 quasars, and 180 contaminants. 

As explained in Section~\ref{sec:convergence}, a sample of 64 plates is considered to be converged. 
We draw 8 test samples (which we name {\it sample 0} to {\it sample 8}) of 64 plates by randomly drawing plates (without repetition). We require that each plate used in a training, a testing or a validation sample be uniquely assigned to that particular sample. 
The training sample of 512 plates is split into 8 training samples of 64 plates in order to have an independent training sample for each of the independent test samples. 

%% file: src/SQUEzE_convergence_tests.tex
\section{Convergence tests}\label{sec:convergence}

Before moving to analysing the performance of SQUEzE we need to make sure that the samples used both to train the classifiers and to evaluate SQUEzE performance are large enough. Ideally, we would use a sample as large as possible for both the training and the testing. However, the larger the sample, the larger the computational requirements. Larger samples require more memory to manage them, and the training is more time-consuming. Also, the preparation of the training sample itself may be demanding. We then need to use samples as small as possible while at the same time ensure that they are converged, i.e., that there will be negligible difference to the performance if we increase the samples. Note that since we are making a choice about the model itself we cannot use the test sample here but we have to use the test samples instead.

To test the convergence of the samples we run SQUEzE on samples of different sizes. For a run of size $i$, we train the model using the training sample of $i$ plates, and then we evaluate the performance of SQUEzE on the validation sample of also $i$ plates. We do this exercice for $i = 2, 4, 8, 16, 32, 64, 128, 256, 512$.

\begin{figure*}
	\centering
	\includegraphics[width=\textwidth]{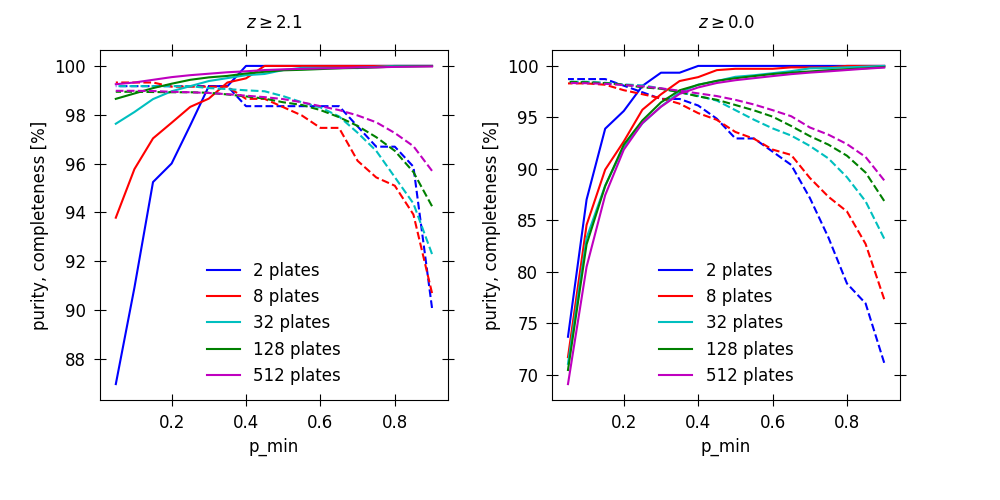}
	\caption{Purity (solid lines) and completeness (dashed lines) as a function of the confidence threshold $p_{\rm min}$, obtained for validation samples of sizes 2, 8, 31, 128, and 512 plates. Right panel shows the results for the entire sample whereas left panel restrict to quasars with $z_{\rm try} \geq 2.1$. We note that on average each plate has 100 quasars and 180 contaminants.}
	\label{fig:convergence_test_vs_p}
\end{figure*}

Figure \ref{fig:convergence_test_vs_p} shows the purity and the completeness of samples of size 2, 8, 32, 128, and 512 plates as a function of the chosen threshold for the confidence parameter, $p_{\rm min}$. The performance is shown for the entire sample, but also for those quasars with $z_{\rm try} \geq 2.1$, which corresponds to quasars where the \lya{} line is present on the optical spectrum. 

\begin{figure}
	\centering
	\includegraphics[width=0.45\textwidth]{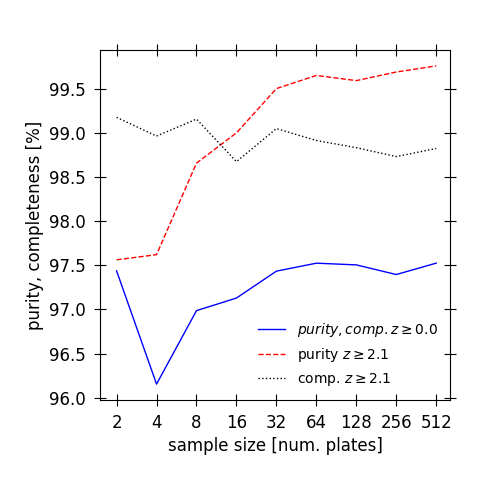}
	\caption{Purity and completeness as a function of sample size. Statistics are given using the $p_{\rm min}$ value where purity and completeness are the same for all the sample. Blue solid line shows the purity and completeness for the entire sample are considered. Purity and completeness are equal because we take the purity and convergence at the crossing point. Red dashed line shows the purity for quasars with $z_{\rm try} \ge 2.1$ only, and black dotted line shows the completeness for quasars with $z_{\rm try} \ge 2.1$. We note that on average each plate has 100 quasars and 180 contaminants. }
	\label{fig:convergence_test}
\end{figure}

We can see that the lines are not overlapping for extreme choices of the confidence threshold. However, if we focus on the confidence threshold where purity and completeness are equal, we can see that the values of purity and completeness have converged. This is shown in more detail in Figure~\ref{fig:convergence_test}.
To quantify the degree of convergence acquired table \ref{ta:convergence_test} gives the maximal variation on the purity and completeness when comparing the results of a given size with those of bigger samples. We conclude that a sample of 64 plates is converged as long as extreme values of the confidence threshold are not considered.

\begin{table*}
	\centering
	\begin{tabular}{ccccc}
	\toprule
	\multirow{2}{*}{sample size} & \multicolumn{2}{c}{maximal variation} & \multicolumn{2}{c}{maximal variation $z \geq 2.1$} \\
	& purity & completeness & purity & completeness \\
	\midrule
	32 & 0.09\% & 0.09\% & 0.26\% & 0.32\%\\
	64 & 0.13\% & 0.13\% & 0.11\% & 0.18\%\\
	128 & 0.11\% & 0.11\% & 0.17\% & 0.10\%\\ 
	256 & 0.13\% & 0.13\% & 0.07\% & 0.09\%\\
	\bottomrule
	\end{tabular}
	\caption{Maximal variation on the purity and completeness when comparing the results of a given size with those of bigger samples.}
	\label{ta:convergence_test}
\end{table*}

%% file: src/SQUEzE_results.tex
\section{SQUEzE performance}\label{sec:results}
We now present the main results regarding SQUEzE performance. As discussed in Section~\ref{sec:convergence}, a sample of 64 plates is considered to be converged. We run SQUEzE on the 8 test samples using the models trained using the 8 training samples of 64 plates. The final value of the metrics is the mean values on the 8 samples, and their errors correspond to the dispersion of the metric among the 8 samples. We would note in advance that all the results which follow presume that the visual inspection catalogue lacks impurities and is complete to quasars targeted spectroscopically. It is evident that this is overwhelmingly the case but, since errors are possible, the performance of SQUEzE may in fact be better than is indicated here. We will investigate this question in a future publication in this series.

\begin{figure}
	\centering
	\includegraphics[width=0.45\textwidth]{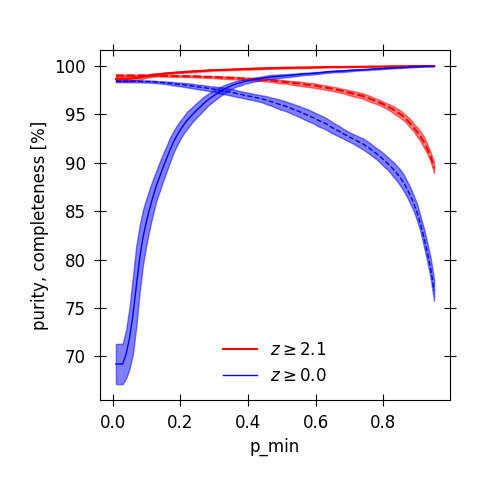}
	\caption{Purity (solid lines) and completeness (dashed lines) as a function of the confidence threshold $p_{\rm min}$, obtained when averaging the results from the 8 test samples (64 plates). Blue lines show the results for the entire sample whereas red lines restrict to quasars with $z_{\rm try} \geq 2.1$. Shaded regions indicate the dispersion found on the 8 test samples. We note that on average each plate has 100 quasars and 180 contaminants.}
	\label{fig:results_p}
\end{figure}

Figure~\ref{fig:results_p} and Table~\ref{ta:results_p} show the purity and completeness as a function of the confidence threshold. If we select a confidence threshold such that the purity and completeness are similar (corresponding to $p_{\rm min} = 0.32$) we obtain a purity of $97.40\pm0.47$ and a completeness of $ 97.46\pm0.33\%$ for the entire sample and a purity of $99.59\pm0.06\%$ and a completeness of $98.81\pm0.13\%$ for quasars with $z_{\rm try} \geq 2.1$. We note that if the goal were to produce, for example, an essentially pure sample of \lya{} quasars (as defined by the BOSS visual inspection) this is currently possible with SQUEzE alone using a threshold of  $p_{\rm min}=0.7$ giving  purity of $99.91\pm0.03\%$. This would come at the cost of sacrificing 1-2\% of the sample (completeness of $97.23\pm0.26\%$).

\begin{table*}
	\centering
	\begin{tabular}{ccccc}
		\toprule
		probability & \multicolumn{2}{c}{purity [\%]} & \multicolumn{2}{c}{completeness [\%]} \\
		threshold  & all & $z \ge 2.1$ & all & $z \ge 2.1$ \\
		\midrule
 0.05 &  $ 71.85 \pm 3.16$&   $98.68 \pm 0.27$&  $98.43 \pm 0.11$&      $99.02 \pm 0.08$\\
 0.10 &  $83.92 \pm 2.39$&   $98.94 \pm 0.16$&  $98.40 \pm 0.11$&      $99.02 \pm 0.08$\\
 0.15 &  $89.59 \pm 1.41$&   $99.20 \pm 0.09$&  $98.28 \pm 0.13$&      $98.99 \pm 0.09$\\
 0.20 &  $93.39 \pm 0.95$&   $99.34 \pm 0.08$&  $98.09 \pm 0.18$&      $98.95 \pm 0.09$\\
 0.25 &  $95.48 \pm 0.75$&   $99.47 \pm 0.08$&  $97.85 \pm 0.24$&      $98.91 \pm 0.10$\\
 0.30 &  $96.95 \pm 0.55$&   $99.57 \pm 0.06$&  $97.61 \pm 0.27$&      $98.86 \pm 0.11$\\
 0.35 &  $97.94 \pm 0.34$&   $99.63 \pm 0.07$&  $97.28 \pm 0.34$&      $98.77 \pm 0.13$\\
 0.40 &  $98.47 \pm 0.24$&   $99.70 \pm 0.09$&  $96.91 \pm 0.34$&      $98.67 \pm 0.11$\\
 0.45 &  $98.78 \pm 0.20$&   $99.75 \pm 0.07$&  $96.52 \pm 0.39$&      $98.56 \pm 0.14$\\
 0.50 &  $98.96 \pm 0.17$&   $99.81 \pm 0.08$&  $95.96 \pm 0.50$&      $98.38 \pm 0.19$\\
 0.55 &  $99.12 \pm 0.14$&   $99.84 \pm 0.07$&  $95.32 \pm 0.57$&      $98.19 \pm 0.22$\\
 0.60 &  $99.26 \pm 0.11$&   $99.87 \pm 0.05$&  $94.53 \pm 0.60$&      $97.95 \pm 0.24$\\
 0.65 &  $99.42 \pm 0.09$&   $99.89 \pm 0.04$&  $93.62 \pm 0.58$&      $97.64 \pm 0.22$\\
 0.70 &  $99.54 \pm 0.09$&   $99.91 \pm 0.03$&  $92.66 \pm 0.49$&      $97.23 \pm 0.26$\\
 0.75 &  $99.64 \pm 0.07$&   $99.93 \pm 0.03$&  $91.71 \pm 0.58$&      $96.75 \pm 0.29$\\
 0.80 &  $99.75 \pm 0.05$&   $99.95 \pm 0.02$&  $90.34 \pm 0.65$&      $96.03 \pm 0.28$\\
 0.85 &  $99.84 \pm 0.05$&   $99.98 \pm 0.01$&  $88.47 \pm 0.83$&      $95.03 \pm 0.39$\\
 0.90 &  $99.93 \pm 0.03$&   $99.99 \pm 0.02$&  $84.85 \pm 0.86$&      $93.19 \pm 0.49$\\
 0.95 &  $99.99 \pm 0.02$&  $100.00 \pm 0.01$&  $76.82 \pm 1.08$&      $89.49 \pm 0.56$\\	
    \bottomrule
	\end{tabular}
	\caption{Evolution of the different metrics for different values of the chosen probability threshold. Value of the metrics is the mean
of the sample 0 to sample 8, and their errors correspond to the standard deviation of the metric among the 8 samples. Each of the samples has a size of 64 plates and uses a different training set, also of 64 plates. We note that on average each plate has 100 quasars and 180 contaminants.}
	\label{ta:results_p}
\end{table*}

\begin{figure*}
	\centering
	\includegraphics[width=\textwidth]{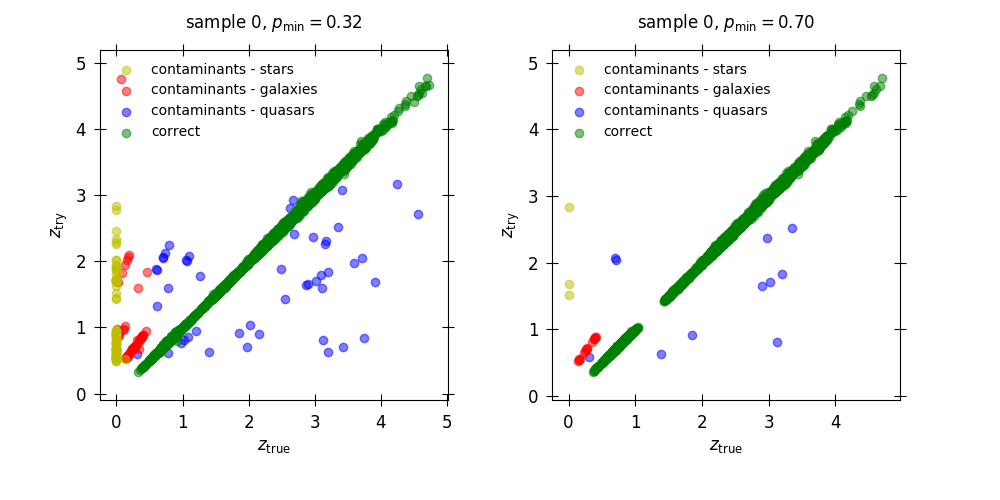}
	\caption{Line confusion plot. For all the objects in the catalogue, $z_{\rm try}$ against $z_{\rm true}$. In this plot, green circles correspond to correct classifications, yellow circles to stellar contaminants, red circles to galactic contaminants, and blue circles to quasar contaminants. Note that this plot signals out the false positives.}
	\label{fig:line_confusion}
\end{figure*}

We now investigate in more detail the failure modes of SQUEzE, with particular focus on the case of purity and completeness equality and a case where purity is prioritised.  Figure \ref{fig:line_confusion} shows the recovered trial redshifts against the true redshifts for sample 0. In this plot, green circles correspond to correct classifications, yellow circles to stellar contaminants, red circles to galactic contaminants, and blue circles to quasar contaminants. Note that this plot singles out the false positives. We present this detailed analysis for sample 0, but the conclusions extracted are extrapolated to the other samples. 
For $p_{\rm min} =  0.32$, we have 6,325 correctly classified quasars at all redshifts, 48 quasars that were assigned a wrong redshift, and 57 galaxies and 91 stars that were classified as quasars. For $p_{\rm min} = 0.7$, we have 6,079 correctly classified quasars at all redshifts, 11 quasars that were assigned a wrong redshift, and 15 galaxies and 3 stars that were classified as quasars. In this sample, there are 6,505 quasars. This results show that the first type of contaminants, stars, can mostly be discarded by raising the confidence threshold. The second type of contaminants, the galaxies, are usually given a higher redshift than that of the visual inspection. 
The quasar contaminants on the sample are quasars which have been classified as such, but have the wrong redshift assigned. This is because SQUEzE misidentified the line responsible for the emission. Particularly for a high confidence threshold, we see that most of these contaminants are given a redshift that is lower to the value reported in the visual inspection.

\subsection{Redshift precision}\label{sec:results_z}
\begin{figure}
	\centering
	\includegraphics[width=0.5\textwidth]{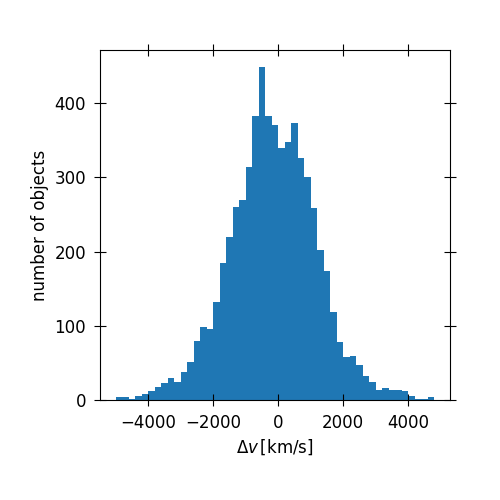}
	\caption{Velocity offset distribution for sample 0 of the quasar candidates that are deemed correct, i.e., those with $\left|z_{\rm try} - z_{\rm true}\right| \le \Delta z_{r} = 0.15$.}
	\label{fig:hist_z}
\end{figure}

In Section~\ref{sec:squeze_general} we stated that a candidate is deemed a true quasar if its trial redshift is closer than $\Delta z_r=0.15$ to the true redshift. However, in most cases the redshift precision is much greater than this. Figure \ref{fig:hist_z} shows the distribution of the offset in velocity, 
\begin{equation}
	\Delta v = \frac{z_{\rm try}-z_{\rm true}}{1+z_{\rm try}} c
\end{equation}	
(where $c$ is the speed of light), of the correct candidates for sample 0 and for $p_{\rm min} =  0.32$.
As we can see most of the values are an order of magnitude smaller than our requirement limit. If we analyse the distribution found in the different test samples, we find the mean redshift offsets and the standard deviations reported in table \ref{ta:hist_z}. The measured standard deviations suggest that our redshift error is typically $\Delta v=1,500$\kms{} (or equivalently $\Delta z = 0.018$). However, at this stage we cannot estimate a {\it per object} redshift error. The errors on the redshift estimation for each of the candidates depend in a non-trivial way on a number of things including (but not limited to) the number of lines detected in the quasar (i.e., the number of times the object is on the list with the correct redshift before step 3), the asymmetry of the peaks, the signal-to-noise of the spectrum and the systematic error associated with emission line transition used. 

\begin{table}
	\centering
	\begin{tabular}{ccc}
	\toprule
	sample & mean [\kms{}] & standard deviation [\kms{}] \\
	\midrule
	0 & \textbf{-138} &  \textbf{1,450} \\
	1 & \textbf{-100} & \textbf{1,461} \\	
	2 & \textbf{-124} & \textbf{1,512} \\
	3 & \textbf{-81} & \textbf{1,495} \\
	4 & \textbf{-67} & \textbf{1,484} \\
	5 & \textbf{-79} & \textbf{1,522} \\
	6 & \textbf{-106} & \textbf{1,507} \\
	7 & \textbf{-95} & \textbf{1,505} \\
	\bottomrule
	\end{tabular}
	\caption{Mean and standard deviation of the redshift distribution for the different test samples for the correctly classified candidates assuming $p_{\rm min}=0.32$.}
	\label{ta:hist_z}
\end{table}


%% file: src/SQUEzE_discussion.tex
\section{Discussion}\label{sec:discussion}
SQUEzE is a machine-learning code that assesses spectra to identify quasars and assign them a redshift. To achieve this, the machine learning step of SQUEzE does not have access to spectra but utilizes high level metrics. Because of this, and because quasars are self-similar, SQUEzE reaches satisfactory purity and completeness even for small training samples. 

The results shown in this paper assume that the visual inspection is always correct. However, there are some known mistakes in the visual inspection catalogue. Based on a preliminary visual re-inspection of a subset of the apparent failures, we believe that the visual inspection catalogue suffers from low levels of impurity and incompleteness (at most 1\%). This may indicate that, in some cases where SQUEzE is in disagreement with the truth table here, SQUEzE may be correct. This will be addressed in follow-up papers of the series.

The results shown here demonstrate SQUEzE performance when training on and applied to BOSS data. A major advantage of SQUEzE is that it utilises only high-level metrics and does not access the entire spectra. This implies that SQUEzE will have a limited capacity to learn spectrograph defects and pipeline artefacts. When applied to surveys with similar signal-to-noise and resolution, such as WEAVE-QSO and DESI, SQUEzE is likely to have similar performance, even using the training set used here. In a follow-up paper of the series we will investigate sensitivity to changes in signal-to-noise and resolution.

In the convergence test (Figure~\ref{fig:convergence_test}) we can see that already a training sample of only 2 plates ($\sim$200 quasars) produces purity and completeness that are above 95\%, when the confidence ratio is chosen such that the purity and completeness are similar. For this purity and completeness quality case, the overall metrics seems to be converged for a sample of size 64 plates ($\sim$6,400) quasars. However, we note that the purity for quasars with $z \geq 2.1$ seems to continue to improve to a near-negligible degree. Give the small proportion of the dataset in question, this may be an indication of rare flaws (or borderline quasars) in the training set.

If we focus on extreme values of the confidence threshold we find that results are converged only on the purity or the completeness, but not on both simultaneously (Figure~\ref{fig:convergence_test_vs_p}). High values of $p_{\rm min}$ ($p_{\rm min}\geq0.7$) give essentially 100\% purity, and this value seems to be converged even for the smallest sample analysed (2 plates). We do not recommend that SQUEzE be used with these extreme choices of threshold probability both because extreme high purity or completeness modes can be achieved within our recommended $0.2\leq p_{\rm min} \leq 0.7$ range, but also because outside of this range training is unnecessarily demanding for convergence.

\begin{figure}
	\centering
	\includegraphics[width=0.5\textwidth]{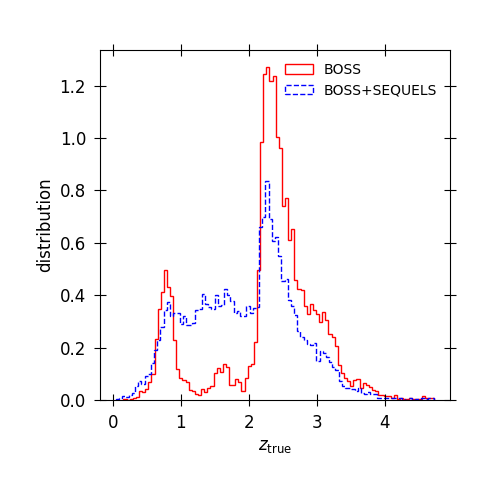}
	\caption{Quasar redshift distribution of test sample 0 before (solid line, with only BOSS plates) and after (dashed line, with plates from BOSS and SEQUELS) being modified (see text for details).}
	\label{fig:quasar_z}
\end{figure}

The redshift distribution of quasars in the samples thus far have not been representative of the real redshift distribution of quasars since the BOSS survey prioritised Lyman-$\alpha$ forest quasars. This leads to somewhat optimistic conclusions with regards the all redshifts tests since Lyman-$\alpha$ forest quasars evidently present a simpler task for SQUEzE.
To order to better assess the performance for a survey intended to target and identify quasars of all redshifts, we replace 32 of the BOSS plates in our sample 0 for 12 SEQUELS\footnote{see, e.g., Section 5 of \cite{Myers+2015}} plates (each of the SEQUELS plates have $\sim$300 quasars, thus the new sample has approximately the same size as prior samples). The difference between the initial quasar redshift distribution and that of the modified samples is shown in Figure~\ref{fig:quasar_z}. 

Taking $p_{\rm min} = 0.32$, this sample returns a purity of 97.08\% and a completeness of  96.22\% for all redshifts.
The performance on this sample yields a completeness $\sim$1\% worse than that of BOSS sample, and a purity $\sim0.5\%$ worse than that yielded by the BOSS sample. This suggests that our results on the BOSS sample may be slightly optimistic compared to a complete sample of quasars, but further investigation on this is required. In a follow-up paper of this series we will analyze in detail the universality of the results presented here.

The last point we want to discuss concerns the redshift precision of the code. As stated in Section~~\ref{sec:results_z}, the typical redshift error is 1,500\kms{}. This redshift error is estimated by measuring the standard deviation of the difference between $z_{\rm try}$ and $z_{\rm true}$. Because this redshift error is rather large for most of the applications, we recommend the user to consider this redshift error as a prior to actual redshift fitters. We note that this error can be reduced to 1,200\kms{} if one computes the mean shift for each assumed line using the correct peaks in the training sample, and then subtracting these mean shift to $z_{\rm try}$. However, there are a number of alternatives to improve the redshift determination. A smaller smoothing of the spectra could be used to refine the position of the peak, but it is unclear how noise will affect this position. The identification of multiple peaks could be also used in several ways to refine these redshifts. Therefore, further studies are required to refine this redshift, and we leave this point for future versions of the code.

We now put SQUEzE in context with the other automatic algorithms present in the literature. We continue the treatment we have used throughout this work and assume that the visual inspection is always correct. We compare the results of SQUEzE with those of the pipeline \citep{Bolton+2012}, and with those of alternative classifier QuasarNet \citep[a classifer based on convolutional neural networks, see][]{Busca+2018}. The purity and completeness for the pipeline is 99.14\% and 94.98\% respectively, whereas for QuasarNet, they report a purity of $99.51\pm0.03$\% and a completeness of $99.52\pm0.03$\%. We also acknowledge the existence of RedRock, an unpublished DESI code that develops the methods used by the BOSS pipeline and that is performing with similar levels of purity and completeness (but no official values are yet reported). We note that these results are for the entire sample. 
If we compare SQUEzE performance with that of the pipeline we can see that we get the same purity if we select a probability threshold of 0.55. By doing so, SQUEzE reaches a slightly higher level of completeness. This is particularly true for quasars with redshift greater than 2.1: we select a probability threshold of 0.15 to achieve a purity of about 99.14\%, obtaining a completeness of 98.99\%.
If we compare the performance of SQUEzE and QuasarNet, we see that QuasarNet is better performing. However, we note that QuasarNet sees the entirety of the spectra, and as such may learn from the mistakes in the visual inspected catalogue. SQUEzE, on the other hand, sees only high level metrics (as opposed to seeing the entirety of the spectra), which makes it more flexible to be used for other surveys.

We stress again that the performance discussed here assume that the visual inspection is always correct. A more detailed analysis of the disagreements between the classifiers (including the visual inspection) will be provided in a follow-up paper. However, we note that this superficial comparison between the classifiers should be enough to establish a classification strategy for new surveys. We recommend that all the available classifiers are used in a combined way. Because the different classifiers have completely different approaches to the classification problem, their agreement will reinforce the confidence of the classification. We currently recommend that all automated analysis are applied to a representative sample that is visually inspected in order to develop a decision tree.

%% file: src/SQUEzE_conclusions.tex
\section{Summary and conclusions}\label{sec:conclusions}

We have presented SQUEzE, an algorithm that follows the human visual inspection to classify quasars and estimate their redshift. We have then tested the performance of this algorithm on visually inspected data from BOSS. We now summarize the key points we have discussed so far.
\begin{enumerate}
	\item SQUEzE mimics the procedure of human visual inspection of the spectra to identify quasars and estimate their redshift. It performs a three step classification of quasars by identifying emission peaks, computing the redshift assuming an emission line, and selecting the correct quasar redshifts from the different possible values of $z_{\rm try}$.
	\item If we assume a confidence threshold of 0.32, we recover a purity of $97.40\pm0.47\%$ and a completeness of $97.46\pm0.33\%$ for the entire sample. This purity and completeness increases to $99.59\pm0.06$ and to $98.81\pm0.13\%$, respectively, if we restrict to objects with $z_{\rm try} \geq 2.1$.
	\item We can achieve virtually 100\% purity if we assume a higher confidence threshold. For instance, for $p_{\rm min}=0.7$ we recover a purity of $ 99.91\pm0.03\%$ and a completeness of $ 97.23\pm0.26\%$ for objects with $z_{\rm try} \geq 2.1$. 
	\item The analysis of a more realistic quasar redshift distribution produced by combining plates from BOSS and SEQUELS suggests that our results are sligthly optimistic: for this analysis, and considering $p_{\rm min}=0.32$, yields a purity and completeness of $ 97.08\%$ and $96.22\%$, respectively. 
	\item We find the standard deviation of our redshift errors to be 1,500\kms{}. We recommend the use of SQUEzE redshift estimates as a prior to some additional redshift tuning. We discuss options to reduce these errors, but further study is required. Preliminary analysis suggest that the errors could be reduced to less than 1,200\kms{}.
\end{enumerate}

 The performance of SQUEzE presented here can be regarded as a conservative assessment since it is assumed that, where SQUEzE disagrees with visual inspection, visual inspection is always correct. We will assess this in future papers in this series. We will also explore the versatility of SQUEzE and its applicability to survey data that differs significantly from the BOSS spectra used here.

%% file: src/acknowledgments.tex
\section*{Acknowledgments}
The authors would like to thank the eBOSS and DESI Collaborations, in particular David Kirkby and Stephen Bailey, for very helpful discussions. This work was supported by the A*MIDEX project (ANR-11-IDEX-0001-02) funded by the ``Investissements d'Avenir'' French Government program, managed by the French National Research Agency (ANR), and by ANR under contract ANR-14-ACHN-0021.

%% file: src/SQUEzE_optimization_step1.tex
\section{Optimization of step 1}\label{sec:opt_step1}

In Section~\ref{sec:squeze_general} we have explained SQUEzE three-steps algorithm in detail. In particular, we have explained that the identification of emission peaks is done using a native peak finder. We have also quoted that we set the width to be 70 pixels, and the significance threshold to 6. In this section we explain the motivation for this choice.

The peak identification is a key component in the algorithm: any quasar that does not have an emission peak identified will be discarded from the sample at this early stage. Because of this, a good completeness is required after this step. However, this also has a huge impact on the later stages of the code as the detected peaks are used both for training the classifiers and classifying, and only the detected peaks are classified.  On the other hand, having too many peaks will increase the number of contaminants, making the classification problem more difficult. 

\begin{figure}
	\centering
	\includegraphics[width=0.45\textwidth]{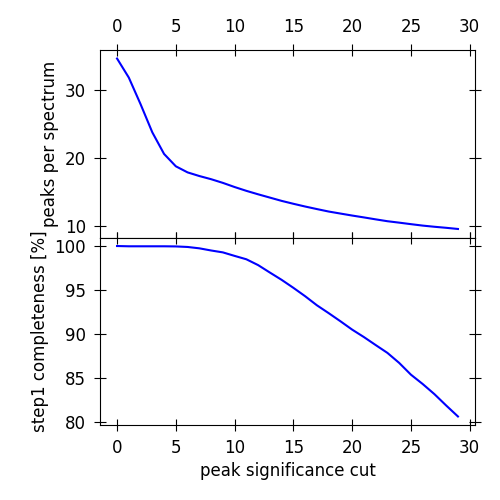}
	\caption{Completeness after step 1 and number of peaks per spectrum as a function of significance threshold.}
	\label{fig:opt_step1}
\end{figure}

We use the training sample of 64 plates to estimate the expected completeness after step 1 and the mean number of peaks per spectrum, as a function of the choice of the significance threshold. The result of this exercise is shown in Figure~\ref{fig:opt_step1}. We find that a significance threshold of 6 offers a good compromise between having a high completeness and a not too high number of peaks per spectrum. For this choice we have around 18 peaks per spectrum, and the completeness after step 1 is 99.88\%.

